\documentclass[superscriptaddress,twocolumn]{revtex4}
\newcommand{\beq}{\begin{equation}}
\newcommand{\eeq}{\end{equation}}
\newcommand{\beqa}{\begin{eqnarray}}
\newcommand{\eeqa}{\end{eqnarray}}

\usepackage{epsfig}
\usepackage{epstopdf}

\begin{document}
\title{Controllable Quantum Switchboard}

\author{D. Kaszlikowski}

\affiliation{Department of Physics, National University of
Singapore, 2 Science Drive 3, Singapore 117542}
\author{L. C. Kwek} \affiliation{Nanyang
Technological University, National Institute of Education, 1,
Nanyang Walk, Singapore 637616}
\author{C. H. Lai}
\affiliation{Department of Physics, National University of
Singapore, 2 Science Drive 3, Singapore 117542}
\author{V. Vedral} \affiliation{\text{The School of Physics and Astronomy,
University of Leeds, Leeds, LS2 9JT, United Kingdom}}

\date{\today}

\begin{abstract}
All quantum information processes inevitably requires the explicit
state preparation of an entangled state. Here we present the
construction of a quantum switchboard which can act both as an
optimal quantum cloning machine and a quantum demultiplexer based
on the preparation of a four-qubit state.
\end{abstract}
\maketitle

Many quantum information processes require the explicit
preparation of specially entangled quantum states. Two-qubit
maximally entangled state often called Bell state, for instance,
form an essential quantum resource needed in quantum
teleportation \cite{bennett}. The preparation of three-qubit
maximally entangled state (like GHZ) could be harnessed for
secure secret sharing \cite{secret}. In one-way quantum
computing, a four-qubit entangled state called cluster state
provides an efficient implementation of a universal quantum gate:
arbitrary single-qubit unitary operation and the CNOT gate \cite{briegel}.

It is interesting to note that entangled states which are used as
a common resource in quantum information processes generally need
not even be maximally entangled at all. As long as the state is
genuinely entangled, quantum computation and communication will
generally be better than the classical counterparts. In
particular, the non-maximally entangled W state has been
experimentally implemented and proposed for controlled quantum
teleportation and secure communication \cite{eibl}.

An essential component of any quantum computation is the ability
to spread quantum information over various parts of quantum
computer. The parts then undergo separate evolutions depending on
the type of the quantum information processing we wish to
implement. Ultimately we need be capable of navigating the
relevant part of the information into a designated output. In a
classical computer this flow of information is achieved through a
controllable switch. Is it possible to design a quantum analogue
for such a device? An added complexity in a quantum switch would
be the requirement that the information flows down many possible
channels coherently as well as the possibility of channeling it
in one selected direction.

Ideally we would like to realize the simplest such a device with
the least number of qubits needed for this purpose. In addition
these qubits will in practice be implemented in a physical system
which will determine the nature of the qubits and couplings
between them. Therefore, when designing our switch we should also
take into account realistic interaction between the qubits, which
severely limits the number of possible Hamiltonians to execute
such a quantum switch. Here we present a possible implementation
of the switch that fulfills of all the above requirements.

Let us consider an interesting four-qubit state described by
\begin{equation} |\psi\rangle =
\frac{1}{\sqrt{3}}\left(|(11)_{12}\rangle|(11)_{34}
\rangle-|(11)_{14}\rangle|(11)_{23}\rangle\right),
\label{state}
\end{equation}
where $|(11)_{ij}\rangle =\frac{1}{\sqrt{2}}(
|0\rangle_i|1\rangle_j-|1\rangle_i|0\rangle_j)$ is the singlet
state. Throughout the paper we use the following notation for the
Bell states $|(ab)\rangle=\sum_{k=0}^3\frac{(-1)^{kb}}{\sqrt
2}|k,k+a\rangle$ with summation modulo $2$.  It turns out that
this state is ideally suited for a quantum switchboard, i.e., a
circuit that can be used to direct the flow of quantum information
in a controllable manner. An interesting property of the
presented switchboard is that in the case of failure the
information is not entirely lost.

The first qubit of the state (\ref{state}) belongs to Alice, the
second one to Bob, the third one to Charlene and the last one to
Dick. Suppose Alice attaches some auxiliary qubit to the first
qubit and perform a joint Bell measurement. Immediately after
getting one of the four possible outcomes, she broadcasts two
(classical) bits of information to Bob and Charlene as it is in the
usual teleportation scheme. At this point, it is not necessary
for Dick to know these two bits of information.

Bob and Charlene can recover the state of the auxiliary qubit with
the fidelity $\frac{5}{6}$ by applying appropriate unitary
transformation based on the knowledge of the broadcast classical
bits. The given state at the beginning does not provide a
universal cloning machine for three copies of the cloned state
\cite{cloning}. Thus, the qubit belonging to Dick is related to
the Alice's auxiliary qubit with the "classical" fidelity
$\frac{1}{3}$, i.e., the fidelity that can be achieved without
prior entanglement. It is interesting to note that Bob and Charlene
possess the optimum fidelity achievable under a symmetric cloning
machine. Dick's fidelity is allowed since there is no limitation
on the production of clones with the fidelity below $\frac{2}{3}$.

Thus the presented protocol behaves like an optimal telecloner
\cite{telecloning}. However, there is still an unused qubit held
by Dick. Depending on Alice's decision regarding to whom she
wishes ultimately to send her auxiliary qubit, say Bob (Charlene)
for instance, she can direct Dick to send his qubit to Charlene
(Bob). As soon as Charlene receives Dick's qubit, he can perform a
Bell measurement on his qubit with Dick's qubit and send the
results of his measurement to Bob. Using the information from
Charlene, Bob can perfectly recover the state of the Alice's
auxiliary qubit.

\begin{figure}[tbp]
\begin{center}
\epsfig{figure=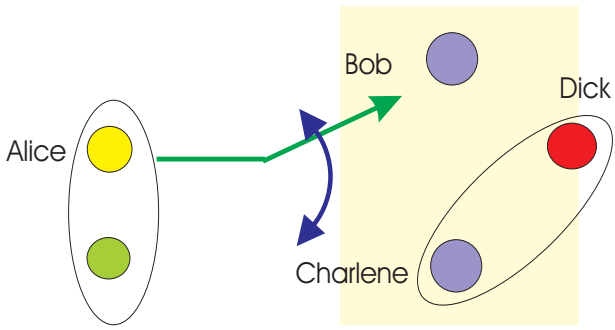, height=.18\textheight,width=0.45\textwidth} 
\caption{Suppose
Alice wishes to send her auxiliary qubit to Bob. She can direct
Dick to send his qubit to Charlene.  Charlene then performs a
Bell measurement on his qubit with Dick's qubit and send the
results of his measurement to Bob. Using the information from
Charlene, Bob can perfectly recover the state of the Alice's
auxiliary qubit. }\label{fig:Tele}
\end{center}
\end{figure}

The situation is entirely symmetric, i.e., Dick can send his qubit
to Bob instead of Charlene with the result that now Charlene can
obtain Alice's auxiliary qubit with perfect fidelity.  In short
the state acts as a quantum switchboard in which Alice can direct
optimal clones to Bob and Charlene or perform perfect quantum
teleportation to Bob or Charlene by utilizing Dick's qubit as in a
quantum demultiplexer. A schematic diagram of this quantum
switchboard protocol is shown in Fig. \ref{fig:Tele}. By
directing Dick's qubit to either Bob (or Charlene), Alice can
effectively transfer the unknown auxiliary qubit to Charlene (or
Bob).  Moreover, she can delay the transfer process to a later
time as long as she has effective control over Dick's qubit.

Incidentally, other shared states may be able to achieve some aspect of our quantum switch but it is difficult to find a state with all desired properties. For instance with the GHZ state,
shared among Alice, Bob and Charlene, one could in principle
provide perfect quantum teleportation to both Bob and Charlene, but
without the additional benefit of an optimal quantum cloner. In
this case, Alice teleclones to both Bob and Charlene with a
classical fidelity of $2/3$. The eventual quantum teleportation
to Bob (or Charlene) is performed with a measurement in the basis
$1/\sqrt{2}(|0\rangle \pm |1\rangle)$. 

It is also interesting to
note that the sheer presence of singlets or dimer-like bonds in
the four-qubit state may make it more robust to certain types of noise. One example would be fluctuating magnetic field or polarization drift, depending on how we implement our qubits. This kind of fault tolerance is absent in the GHZ state. 

Let us now prove the above statements. It is
convenient to write the state $|\psi\rangle$ in the following way
\begin{eqnarray}
&&|\psi\rangle = \frac{1}{2\sqrt{3}}(3|(11)_{12}
\rangle|(11)_{34}\rangle +|(10)_{12}\rangle |(10)_{34}\rangle+\nonumber\\
&&|(01)_{12}\rangle |(01)_{34}\rangle-|(00)_{12}\rangle
|(00)_{34}\rangle).
\end{eqnarray}
We can immediately see that the state shared by Alice and Bob is
the Werner state with $\frac{1}{3}$ of noise. Taking into account
that the fidelity of teleportation for the Werner state with the
noise fraction $1-p$ is given by $\frac{p+1}{2}$
\cite{telewerner} we see that the fidelity of Bob's qubit is
$\frac{5}{6}$. It can be checked that the state between Alice and
Dick is the Werner state that is an equal mixture of the three
Bell states $|(10)\rangle, |(01)\rangle, |(00)\rangle$. Thus
Dick's clone of Alice's auxiliary qubit has the fidelity
$\frac{1}{3}$, which is the fidelity achievable classically.

The state $|\psi\rangle$ is symmetric with respect to Bob and
Charlene
\begin{eqnarray}
&&|\psi\rangle = \frac{1}{2\sqrt{3}}(3|(11)_{13}
\rangle|(11)_{24}\rangle +|(10)_{13}\rangle |(10)_{24}\rangle+\nonumber\\
&&|(01)_{13}\rangle |(01)_{24}\rangle-|(00)_{13}\rangle
|(00)_{24}\rangle).
\end{eqnarray}
therefore Charlene's clone has the same fidelity as Bob's one.

Let us now write the state $|\psi\rangle$ together with the
Alice's auxiliary qubit $|\alpha\rangle$ (particle with $0$
index) in the form suitable for further analysis. To focus our
attention we consider the scenario where Dick sends his qubit to
Charlene. We have
\begin{eqnarray}
&&|\alpha\rangle|\psi\rangle=\frac{1}{4\sqrt{3}}\sum_{k,l,m,n=0}^1\lambda_{kl}|(mn)_{01}\rangle\otimes\nonumber\\
&&\otimes ~U_{mn,kl}|\alpha\rangle|(kl)_{34}\rangle, \label{tele}
\end{eqnarray}
where
$\lambda_{11}=3,\lambda_{01}=\lambda_{10}=1,\lambda_{00}=-1$ and
$U_{mn,kl}$ is a usual unitary transformation that appears in the
process of teleportation with the Bell state $|(kl)\rangle$ and
with the outcome of Bell measurement $(mn)$. For instance,
$U_{01,11}=\sigma_x$.

Suppose now that the outcome of Alice's measurement is $(mn)$.
The collapsed state $|\chi_{mn}\rangle$ shared by Bob, Charlene
and Dick is
\begin{eqnarray}
|\chi_{mn}\rangle =
\sum_{k,l=0}^1\frac{\lambda_{kl}}{2\sqrt{3}}U_{mn,kl}|\alpha\rangle|(kl)_{34}\rangle.
\end{eqnarray}

Therefore, Bob's state $\rho_{mn}$ reads
\begin{eqnarray}
&&\rho_{mn} = \frac{1}{12}\sum_{k,l=0}^1
|\lambda_{kl}|^2U_{mn,kl}|\alpha\rangle\langle\alpha|U_{mn,kl}^{\dagger}.
\end{eqnarray}
After receiving two bits $(mn)$ of classical information from
Alice, Bob can recover Alice's state with the fidelity
$\frac{5}{6}$ as mentioned before. However, when Charlene performs
the Bell measurement on his and Dick's qubit, obtains the result
$(kl)$ and sends it to Bob, Bob receives the state
\begin{eqnarray}
&& \frac{\lambda_{kl}}{|\lambda_{kl}|} U_{mn,kl}|\alpha\rangle,
\end{eqnarray}
which he can transform back to the state $|\alpha\rangle$ by
applying the inverse unitary transformation $U_{mn,kl}^{\dagger}$.

The symmetry of the state $|\psi\rangle$ allows us to repeat the
same argument for the case in which Alice decides to send her
qubit to Bob so that now Charlene can obtain the state
$|\alpha\rangle$ with perfect fidelity. It is interesting to note
that the relative phase between the components of the state
$|\psi\rangle$ is crucial for desired functionality. Other phase
choices or, for that matter, the complete lack of coherence, will
not give us the same quantum switch.





Finally we emphasise that our quantum switch state is a ground state, albeit degenerate, of the 
Majundar-Ghosh (MG) model \cite{Majumdar}. This
spin chain belongs to a class of many- body Hamiltonians that
provide a good qualitative account of materials like
Cu$_2$(C$_5$H$_{12}$N$_2$)$_2$Cl$_4$, CuGeO$_3$ and
YCuO$_{2.5}$\cite{azuma}.  Therefore our quantum switch is very realistic since it 
may already exist in some solid state systems. MG is essentially a one-dimensional
quantum spin chain with nearest- and next-nearest-neighbor
exchange interactions described by the the Hamiltonian \beq
H_{MG} = J \sum_{i = 1}^N \left( 2 \vec{S}_i \vec{S}_{i+1} +
\alpha \vec{S}_i \vec{S}_{i+2} \right),\eeq where $J>0$ and $N$
is the number of sites in the one-dimensional lattice with
periodic boundary condition. The Hamiltonian is exactly solvable
for $\alpha =1$ and has a quantum phase transition from an ordered
phase to a disordered spin-liquid-like phase as $\alpha$ varies
from zero to some critical value $\alpha_{crit} =
0.482$\cite{Okamoto}.

At $\alpha =1$ and for an even $N$, there is a two-fold degenerate
ground state subspace spanned by two dimer configurations
\begin{eqnarray}
&&|(11)_{12}\rangle|(11)_{34}\rangle\dots |(11)_{(N-1)N}\rangle\nonumber\\
&&|(11)_{23}\rangle|(11)_{45}\rangle\dots |(11)_{N1}\rangle,
\end{eqnarray}
superposition of which, for $N=4$, gives us the state
$|\psi\rangle$.

In conclusion, we have provided a quantum switchboard which could
act both as an optimal quantum cloning machine or a quantum
demultiplexer. Moreover, we also note that it is possible to
extend the switchboard to higher spins and higher dimensional
spaces as long as we have a configuration of dimer-like
neighboring bonds. We also note that apart from spin chains, it is possible that
the four-qubit state considered in this paper could also be
created from multi-photon entangled states generated with
spontaneous parametric down conversion and linear optics
apparatus.

\section{Acknowledgment}

D.K. would like to thank Alastair Kay and Ravishankar Ramanathan for useful discussions.

\end{document}